\documentclass[aps,showkeys,showpacs,12pt,preprintnumbers,nofootinbib,superscriptaddress]{revtex4}
\usepackage[letterpaper]{geometry}                
\usepackage{graphicx}
\usepackage{amssymb,amsmath}
\usepackage{epstopdf}
		\def\n{\nu}

\def\L{\Lambda}

\def\be{\begin{equation}}
\def\ee{\end{equation}}

\begin{document}

\preprint{BRX-TH 611}
\preprint{CALT 68-2745}

\title{A note on positive energy of topologically massive gravity }

\author{S. Deser \\
{\it Physics Department, Brandeis University, Waltham, MA}\\
{\it Lauritsen Laboratory, California Institute of Technology, Pasadena CA}\\}


\date{\today}                                           

\begin{abstract}
I review how ``classical SUGRA" embeddability establishes positive energy E for $D=3$ topologically massive gravity (TMG), with or without a cosmological term, a procedure familiar from $D=4$ Einstein gravity (GR). It also provides explicit expressions for E. In contrast to GR, E is not manifestly positive, due to the peculiar two-term nature of TMG.    
\end{abstract}

\pacs{0465.+e}

\maketitle

\section{Introduction}

The cosmological (CTMG) extension [1] of topologically massive gravity (TMG) [2], 
is undergoing explosive interest (for one recent entry into the already enormous literature, see [3]). A central aspect of the problem concerns TMG's stability properties: 
does it share the energy positivity of $D=4$ Einstein gravity (GR)? My purpose is first 
to review the original positive energy derivations of [1,4], using the classical supergravity (SUGRA) embedding so effective in establishing positivity of GR, and second, also as 
in GR, to provide an explicit expression for the energy. The resulting, physically gratifying, positivity outcome is tempered by the fact that, unlike in GR, this expression 
is not manifestly positive, due to the peculiar nature of TMG as a two-term system. While desirable, such explicit corroboration of the positivity theorem is not strictly necessary.

\section{SUGRA extension}

Historically, the original proof of GR positivity came from full quantum SUGRA [5], 
in the guise of a formally positive expression for the energy operator. It was later adapted to GR by considering the classical, $\hbar \rightarrow 0$, limit of its matrix elements in no-gravitino external states [6]. The subsequent celebrated, seemingly SUGRA-free 
spinorial derivation [7] inspired a ``classical SUGRA" approach [8,9], with no quantum  
fields: the gravitino is just a ``catalyst" here. This procedure is applicable to any supersymmetrizable models, including, as shown in [1,4], TMG and CTMG; their consequent energy positivity was duly noted there. We will reproduce those results, 
then obtain the (C)TMG counterparts of the explicit GR energy formula.    

The original SUSY, and later SUGRA quantum  theorems involved physical fermion fields. Their basis was SUSY's ``square root" nature, expressed by the simple formula
\begin{equation}
H = {1\over 2} tr(Q^2) > 0
\end{equation}
in gravitational and Planck units. It states that the Hamiltonian is proportional to the square of the (Majorana) spinorial charge, hence is non-negative for positive Hilbert space metric, and that it vanishes only at vacuum. Hilbert space positivity is guaranteed by a ghost-free bosonic (consequently also fermionic) spectrum, and is clearly an essential ingredient: For example, in any physical theory, flipping the sign of the action makes both the metric and the energy negative. For GR, the kinetic energy of free gravitons is positive: they are normal massless spin 2 fields. For TMG, things are a bit more complicated. Positivity of its (now massive) ``gravitons" was established in [1], but  required that the Einstein part of the 2-term TMG action have the opposite sign than the normal one in GR. While unexceptionable in TMG, this sign requirement leads to a possible paradox in the CTMG extension, because the latter permits an explicit solution, the BTZ black hole [10], that seems to have negative mass with this sign choice; we shall return (briefly) to this, and to another purely CTMG, issue below. 

\section{TMG Energy}

We begin with a summary of classical SUGRA, as applied to (C)TMG, following the conventions of [9]. Energy is entirely concerned with the bosonic and fermionic initial time constraints (rather than with their evolution equations), 
\begin{align}
{\cal G}^{0\nu} &\equiv G^{0\nu} + \mu^{-1} C^{0\nu} + \Lambda g^{0\nu} \nonumber = 0 \nonumber \\
R^0 &\equiv  f^0 - {1 \over 2 \mu} \epsilon^{ij} \gamma_\nu \gamma_j \bar{D}_i f^\nu = 0 \nonumber \\
C^{\mu\nu} &\equiv \epsilon^{\mu\alpha\beta}D_\alpha \bar{R}_\nu{}^\beta,~~~f^\n \equiv \epsilon^{\nu\alpha\beta}\bar{D}_\alpha \psi_\beta \nonumber \\
\bar{R}_\beta{}^\nu &\equiv R_\beta{}^\nu - {1\over4} \delta_\beta^\nu R.
\end{align}
Here the Einstein tensor $G^{\mu \nu}$ is accompanied by the Cotton tensor $C^{\mu \nu}$, as well 
as by a negative (or zero) AdS cosmological term. The (linear) fermionic operator,  
\begin{equation}
R_L^0 \equiv \epsilon^{ij} \partial_j \left[ \psi_j - {1 \over 2 \mu} \gamma_\nu \gamma_j f^\nu \right],
\end{equation}
is also a two-term sum: ordinary ($D=3$) Rarita-Schwinger plus ``fermionic Chern-Simons". The above system is invariant under the usual local SUSY transformations,
\begin{align}
\delta \psi_\mu &= 2 \bar{D}_\mu \alpha, ~~~\delta e_{\mu a} = i \bar{\alpha} \gamma_a \psi_\mu \\
\bar{D}_\mu \alpha &\equiv \left(D_\mu + {1\over 2} \sqrt{-\Lambda} \gamma_\mu\right)\alpha \nonumber.
\end{align}

Note that a cosmological term shifts the usual covariant derivatives to ``cosmological" ones, whose commutator is ($R + \Lambda g$) and vanishes in AdS; it also explains why only AdS 
sign is allowed by SUGRA. In this classical context, things are much simpler than 
in full quantum SUGRA: First, we need not keep any nonlinear terms in the fermion 
field since it will disappear  after the appropriate variations are performed. Second, 
all quantities are commuting, Grassmann-, rather than Clifford-, valued. The relevant relations are
\begin{align}
Q =  \oint dS_i \epsilon^{ij} &\left[  \psi_j - {1 \over 2 \mu} \gamma_\nu \gamma_j f^\nu \right] , \\
\delta_2 Q(\alpha_1) = \left[\alpha_1 Q, \alpha_2 Q \right] &= - \bar{\alpha}_1 \gamma^\mu \alpha_2 P_\mu \nonumber
\end{align}
here the supercharge Q is defined, as usual in gauge theories, as the spatial integral 
of the linear part of the total (supercurrent $R^0$) constraint, namely of $R^0_L$ , and we must vary it twice to obtain the energy according to the fundamental SUSY commutator.  [The variation parameters here mean their (constant) asymptotic values.] These relations are the classical equivalent to (1), as explained earlier. They display the positive energy of TMG, up to the usual fine print regarding asymptotic behavior of solutions. However, the above results are rather abstract, as compared to the famous [7] explicit manifestly positive E expression in $D=4$ GR, 
\begin{equation}
E_{GR} = \int d^3 x \left[(\nabla \epsilon)^2 + (\bar{\epsilon} \gamma_\mu \epsilon) G^{0\mu}\right].
\end{equation}
This result also follows directly  in the present framework. Here, the $\epsilon$ parameter is the special version of the $\alpha$ in (4) subject to the gauge choice that it obey the spatial Dirac equation $\gamma^i D_i \epsilon=0$. [The value of E, as against its form, is gauge-independent by construction: It has the same value in any gauge, as detailed in [11], but its positivity is cleanly displayed in this one.] Obviously if the sources obey the dominant energy condition, essentially $|T^{00}| > |T^{0i}|$, or are absent, then the second integral is non-negative, or zero, and the first one is manifestly non-negative. Further, the energy vanishes only if $\nabla \epsilon=0$ and there are no (positive) sources; but this implies flat space ( there are no constant curved space spinors). We now obtain the equivalent of (6) for TMG; it can be derived straightforwardly by varying $R^0_L$, exactly as was done for GR in [8, 9]: one converts the spatial integral of the linearized supercurrent, a total divergence, into an integral on its boundary at infinity, performs the indicated variations, and goes back to volume form. We merely cite the result of these slightly tedious gamma-matrix, variation, and integration operations,
\begin{equation}
E_{TMG} = \int d^2 x \epsilon^{ij} \left[ (\bar{D}_i \bar{\alpha})(\bar{D}_j \alpha) - {1\over 2 \mu} \bar{D}_i (\bar{\alpha} \gamma_\mu \alpha) \bar{R}_j{}^\mu \right] + \int d^2 r \bar{\alpha} \gamma_\mu \alpha {\cal G}^{0 \mu}.
\end{equation}
This is our new, explicit 3-term counterpart of (6); it is obviously not manifestly positive, despite the third term's similarity to that in (6). The problem lies in the first two; indeed, we are seeing here again the fact that TMG's properties are entirely ``cooperative": only the sum of its two terms and of their variations contains dynamics, whereas either one alone is entirely dull: pure Einstein means flat (since Ricci-flatness is flatness in $D=3$), while Chern-Simons alone means conformally flat (since Cotton is the $D=3$ Weyl tensor). In either of these truncations, the energy (properly) vanishes, in agreement with the corresponding truncations of (2). So, just as the sum of both actions is required for dynamics, one must also somehow use both contributions to display positivity of the sum of the first two terms of (7). While we have not succeeded (nor tried hard) in this task, the classical SUGRA positivity theorem suffices to guarantee its outcome. [Gratifyingly,   E was indeed positive for all specific solution configurations considered in [12].]

\section{Conclusions}

General properties of locally supersymmetrizable systems ensure that energy of 
(C)TMG is necessarily non-negative, when the proper, ``wrong", sign of its Einstein 
part is chosen to ensure positive energy free excitations. These results are in no 
way compromised by our explicit expression's lacking the manifest positivity of (6), though it will be gratifying to find a gauge displaying this . 

Finally, I turn briefly to two separate, unsettled, aspects of CTMG energy: The first has 
to do with the $D=3$ BTZ black hole, a locally AdS solution of AdS $D=3$ GR [10] whose energy there is positive only if one uses the opposite to ours, ``right", sign Einstein action. This poses no paradox for the nondynamical, pure GR: its action's sign is otherwise totally immaterial, since there are no bulk degrees of freedom. A problem only arises in CTMG, which BTZ, like any Einstein space, automatically also solves: a constant curvature's (covariant) curl vanishes automatically. Hence the dilemma that choosing 
the Einstein term's sign gives up either positivity of the bulk, or of the BTZ, solutions. Possible ways out, involving, e.g., superselection rules, are discussed in [3,13], to which we refer for details; the matter is certainly not yet settled. A very different problem,raised in [14], concerns the role of boundary conditions for linearized solutions of CTMG at a special point, $\mu^2= -\L$, of parameter space and the possibility of their instability; it is treated in careful detail in [3]. This is a difficult technical issue involving the-- always tricky-- evolution of systems due to the pathology of AdS, and the choice of asymptotic decay rates, which must be decided on a physical basis, rather than a priori. The point here is roughly that the constraint equations lead to a slow(er), possibly unacceptable, decay of the energy constraint variable V irrespective of the asymptotics of the linearized bulk modes h: The $\nabla^2 V \sim |\nabla h|^2$ Poisson-like constraint equation always implies
monopole falloff for V. [The analogy in $D=4$ GR is that, however rapidly graviton  excitations die off at spatial infinity, the energy constraint metric component they induce only decays as $1/r$: Demanding $1/r^2$ falloff of the entire metric would forbid any solutions. Only $E=0$, namely flat space is consistent with too fast a required decay.] 
In any case, this possible CTMG stability issue, while important, is confined to the 
above special parameter point.

\section{Acknowledgement}
I thank my CTMG collaborators, S Carlip, A Waldron and D Wise, as well as E Sezgin, for helpful discussions. This work was supported by NSF PHY 07-57190 and DOE DE-FG02-92ER40701 grants to Brandeis and Caltech respectively.

\section{References}

1. S. Deser, in ``Quantum theory of Gravity", ed S.M. Christensen (A. Hilger, London 1984).

2. S. Deser, R. Jackiw \& S. Templeton, {\it Ann. Phys.}, {\bf 140}, 372, (1982); {\it Phys. Rev. Lett.} {\bf 48}, 975, (1982).

3. S. Carlip, arXiv:0906.2384 [hep-th].

4. S. Deser and J.H. Kay, {\it Phys. Lett.}, {\bf B120}, 97, (1983). 

5. S. Deser and C. Teitelboim, {\it Phys. Rev. Lett.}, {\bf 39}, 249 (1977).

6. M. Grisaru, {\it Phys. Lett.}, {\bf B73}, 207 (1978).

7. E. Witten, {\it Comm. Math. Phys.}, {\bf 80}, 381 (1981); T. Parker \& C.H. Taubes, {\it ibid}, {\bf 84}, 223 (1982).

8. G.T. Horowitz and A. Strominger, {\it Phys. Rev.}, {\bf D27}, 2793 (1983); C.M. Hull, {\it Comm. Math. Phys.}, {\bf 90}, 545 (1983).

9. S. Deser, {\it Phys. Rev.}, {\bf D27}, 2805 (1983); D. Boulware, S. Deser and K.S. Stelle in ``Quantum Theory of Fields", eds Batalin et. al. (A Hilger, Bristol 1987).

10. M. Banados, C. Teitelboim and J. Zanelli, {\it Phys. Rev. Lett.}, {\bf 69}, 1849 (1992).   

11. S. Deser, {\it Phys. Rev.}, {\bf D30}, 2805 (1984).

12. E. Sezgin and Y. Tanii, arXiv:0905.3779 [hep-th].

13. S. Carlip, S. Deser, A. Waldron and D. Wise, arXiv:0803.3998, 0807.0486 [hep-th].

14. A. Maloney, W. Song, and A. Strominger, arXiv:0903.4573 [hep-th].



\end{document}